\documentclass[preprint,preprintnumbers,amsmath,amssymb,superscriptaddress]{revtex4}

\usepackage{graphicx}
\usepackage{dcolumn}
\usepackage{bm}

\begin{document}


\title{First observation of \textit{spin}-helical Dirac fermions and topological phases in undoped and doped Bi$_2$Te$_3$ revealed by spin-ARPES spectroscopy}

\author{D. Hsieh}
\affiliation{Joseph Henry Laboratories of Physics, Princeton
University, Princeton, NJ 08544, USA}
\author{Y. Xia}
\affiliation{Joseph Henry Laboratories of Physics, Princeton
University, Princeton, NJ 08544, USA}
\author{D. Qian}
\affiliation{Joseph Henry Laboratories of Physics, Princeton
University, Princeton, NJ 08544, USA}
\author{L. Wray}
\affiliation{Joseph Henry Laboratories of Physics, Princeton
University, Princeton, NJ 08544, USA}
\author{J. H. Dil}
\affiliation{Swiss Light Source, Paul Scherrer Institute, CH-5232,
Villigen, Switzerland} \affiliation{Physik-Institut, Universit\"{a}t
Z\"{u}rich-Irchel, 8057 Z\"{u}rich, Switzerland}
\author{F. Meier}
\affiliation{Swiss Light Source, Paul Scherrer Institute, CH-5232,
Villigen, Switzerland} \affiliation{Physik-Institut, Universit\"{a}t
Z\"{u}rich-Irchel, 8057 Z\"{u}rich, Switzerland}
\author{J. Osterwalder}
\affiliation{Physik-Institut, Universit\"{a}t Z\"{u}rich-Irchel,
8057 Z\"{u}rich, Switzerland}
\author{L. Patthey}
\affiliation{Swiss Light Source, Paul Scherrer Institute, CH-5232,
Villigen, Switzerland}
\author{A. V. Fedorov}
\affiliation{Advanced Light Source, Lawrence Berkeley National
Laboratory, Berkeley, CA 94720, USA}
\author{H. Lin}
\affiliation{Department of Physics, Northeastern University, Boston,
MA 02115, USA}
\author{A. Bansil}
\affiliation{Department of Physics, Northeastern University, Boston,
MA 02115, USA}
\author{D. Grauer}
\affiliation{Department of Chemistry, Princeton University,
Princeton, NJ 08544, USA}
\author{R. J. Cava}
\affiliation{Department of Chemistry, Princeton University,
Princeton, NJ 08544, USA}
\author{Y. S. Hor}
\affiliation{Department of Chemistry, Princeton University,
Princeton, NJ 08544, USA}
\author{M. Z. Hasan}
\affiliation{Joseph Henry Laboratories of Physics, Princeton
University, Princeton, NJ 08544, USA} \affiliation{Princeton Center
for Complex Materials, Princeton University, Princeton, NJ 08544,
USA}

\date{\today}

\maketitle


\textbf{Electron systems that possess light-like dispersion relations or the conical Dirac spectrum, such as graphene \cite{Novoselov(Graphene), Checkelsky} and bismuth \cite{Li, Fuseya}, have recently been shown to harbor unusual collective states in high magnetic fields. Such states are possible because their light-like electrons come in spin pairs that are chiral, which means that their direction of propagation is tied to a quantity called pseudospin that describes their location in the crystal lattice. An emerging direction in quantum materials research is the manipulation of atomic spin-orbit coupling to simulate the effect of a spin dependent magnetic field, in attempt to realize novel spin phases of matter \cite{Day, Koralek}. This effect has been proposed to realize systems consisting of unpaired Dirac cones that are helical, meaning their direction of propagation is tied to the electron spin itself, which are forbidden to exist in ordinary Dirac materials such as graphene or bismuth. Such unpaired spin-helical Dirac cones are predicted to exist at the edges of certain types of topologically ordered materials \cite{Fu:STI2, Ran_dislocation} and support exotic collective states that exhibit non-Maxwell electromagnetic behavior \cite{Essin,Franz,Qi_monopole, zhang_bete}, charge fractionalization \cite{Moore_exciton} and topological proximity effects when interfaced with conventional electron systems \cite{Fu_Majorana2,Akhmerov}.
The experimental existence of topological order cannot be determined without spin-resolved measurements \cite{Fu:STI2,Moore:STI1,Essin}. Here we report a spin- and angle-resolved photoemission study of the hexagonal surface of the Bi$_{2-x}$Mn$_x$Te$_3$ series, which is found to exhibit a single helical Dirac cone that is fully spin-polarized. Our observations of a gap in the bulk spin-degenerate band and a spin-resolved surface Dirac node close to the chemical potential show that the low energy dynamics of Bi$_2$Te$_3$ is dominated by the unpaired spin-helical Dirac modes that are entangled over the macroscopic dimensions of the crystal. Our spin-texture measurements prove the existence of a rare topological phase in this materials class for the first time, and makes it suitable for novel 2D Dirac spin device applications beyond the chiral variety or traditional graphene.}

Three-dimensional topologically ordered materials represent a new phase of quantum matter where a combination of relativistic and quantum entanglement effects dominate \cite{Day}. They were recently proposed \cite{Fu:STI2,Moore:STI1} and shortly afterwards first discovered in the Bi$_{1-x}$Sb$_x$ alloy material \cite{Hsieh_Nature}. In these systems, spin-orbit coupling gives rise to non-current carrying electronic states in the bulk and robust conducting states along the edges. In contrast to graphene, which has four Dirac cones (2 doubly degenerate cones at the K and K$'$ points in momentum space) \cite{Checkelsky}, the remarkable property of topological edge states is that their dispersion is characterized by an odd number of spin-polarized Dirac cones. Therefore, it would be impossible to prove the existence of topological order in materials without probing the spin degrees of freedom as demonstrated recently by Hsieh et.al.\cite{Hsieh_Science}. Odd spin edge metals are expected to exhibit a host of unconventional properties including a fractional (half-integer) quantum Hall effect \cite{Fu:STI2} and immunity to Anderson localization \cite{Fu:STI2,Schnyder}. The most exotic physics, however, may occur at the interface between a topologically ordered material and an ordinary ferromagnet or superconductor, where electromagnetic responses that defy Maxwell's equations \cite{Qi_monopole,Essin,Franz} and excitations that obey non-Abelian statistics \cite{Fu_Majorana2,Akhmerov} have been predicted.

The key to testing such experimental proposals is to find the most elementary form of helical Dirac spin states that are fully spin polarized. Despite an active search no spin helical Dirac cones have been experimentally observed in Bi$_{1-x}$Sb$_x$ \cite{Hsieh_Nature, Hsieh_Science}, Bi$_2$Se$_3$\cite{Xia} or Bi$_2$Te$_3$\cite{Noh}. 
At present no spin-resolved surface band calculation on Bi$_2$Te$_3$ exists and spin-texture is theoretically unknown. Here we report both a conventional and spin-resolved ARPES investigation of single crystals of Bi$_2$Te$_3$ and Bi$_{2-x}$Mn$_{x}$Te$_3$. Remarkably, a single helical Dirac cone is realized on the (111) surface, with fully polarized spins rotating by 360$^{\circ}$ around a single Fermi surface.
Although we observe a time dependent band bending of the surface states, a long standing problem with Bi$_{2}$Te$_3$ \cite{Noh}, we show a method that the characteristic relaxation rate can be systematically controlled by adding small amounts of Mn, and the spin Dirac node can be fairly stabilized near the Fermi level $E_F$. Using a synchrotron light source with a tunable photon energy ($h\nu$), we show that the bulk-like states of Bi$_2$Te$_3$ are insulating with the valence band maximum lying around 100 meV below $E_F$, consistent with the long-theorized six-peak dispersion model. Fully gapped bulk-like states together with a spin-polarized Dirac node lying at a very low energy makes Bi$_2$Te$_3$ and slightly Mn-doped Bi$_{2-x}$Mn$_{x}$Te$_3$ suitable for providing 2D helical spin-textured Dirac fermions for spin device integration.

\begin{figure}
\includegraphics[scale=0.55,clip=true, viewport=-0.0in 0in 10.5in 8.0in]{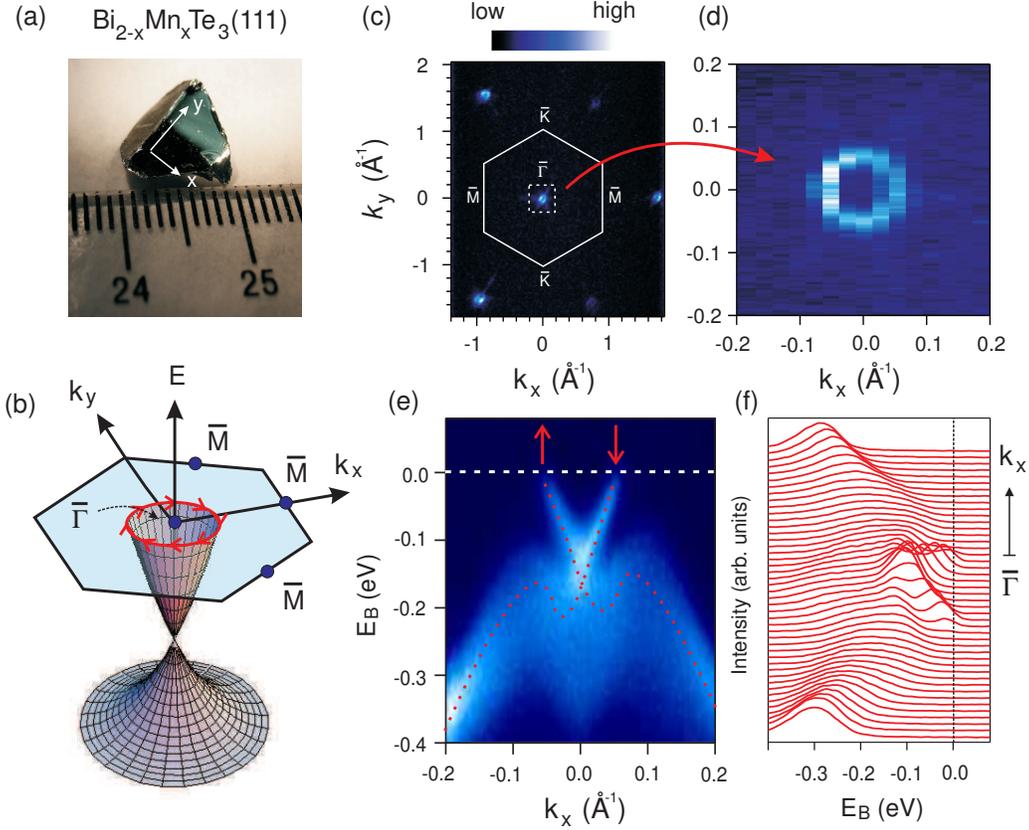}
\caption{\label{fig:STI}. \textbf{A single massless Dirac cone on the surface of Bi$_2$Te$_3$.} (a) Photograph of the cleaved (111) surface of Bi$_2$Te$_3$. (b) Schematic of the (111) surface Brillouin zone with the four time-reversal-invariant momenta ($\bar{\Gamma}$,3$\times$\={M}) marked by blue circles. A single spin-polarized Fermi surface enclosing $\bar{\Gamma}$ that arises from a spin-textured Dirac cone is the direct signature of topological order. (c) Constant energy ARPES intensity map of Bi$_2$Te$_3$ collected at $E_F$ using $h\nu$ = 100 eV. (d) High resolution view of the Fermi surface, using $h\nu$ = 35 eV, over a momentum space window demarcated by the white box in (c). (e) ARPES intensity map of the gapless surface state bands. The red dotted lines are guides to the eye showing the Dirac dispersion. (f) Energy distribution curves of the data shown in (e).}
\end{figure}

High resolution ARPES data were taken at Beamlines 12.0.1 and 7.0.1 of the Advanced Light Source in Lawrence Berkeley National Laboratory.
Spin-resolved ARPES data were taken with the COPHEE spectrometer \cite{HoeschFSmap} at the SIS beamline of the Swiss Light Source using horizontally polarized 22 eV photons. At a pass energy of 3 eV, the energy resolution for spin-detection was set at around 30 meV. The COPHEE spectrometer is capable of measuring all three spatial components of the spin polarization vector for any point in reciprocal space from which a spin-resolved band structure is directly imaged. Data were taken from single crystal Bi$_{2-x}$Mn$_{x}$Te$_3$ cleaved along its (111) surface (Fig.~\ref{fig:STI}(a))
in ultra high vacuum (UHV) at pressures better than $5\times10^{-11}$ torr and maintained at a temperature of 15 K.
Single crystals of Bi$_{2-x}$Mn$_x$Te$_3$ were grown by melting stoichiometric mixtures of elemental Bi (99.999 \%), Te (99.999 \%) and Mn (99.95 \%) at 800$^{\circ}$C overnight in a sealed vacuum quartz tube. The crystalline sample was cooled over a period of two days to 550$^{\circ}$C, and maintained at the temperature for 5 days. It was then furnace cooled to room temperature. The crystal structure of Bi$_{2-x}$Mn$_x$Te$_3$ is rhombohedral with space group $D^5_{3d}$($R\bar{3}m$). It can be visualized as a stack of hexagonal atomic layers, each consisting of only Bi/Mn or Te. Five atomic layers are stacked in a close-packed fcc fashion along the [111] direction in order Te(1)-Bi-Te(2)-Bi-Te(1), in a quintuple layer, and cleavage takes place naturally between such layers. The theoretical calculations were performed with the LAPW method in slab geometry using the WIEN2K package \cite{Wien}. GGA of Perdew, Burke, and Ernzerhof \cite{Ernzerhof} was used to describe the exchange-correlation potential. Spin-orbit coupling was included as a second variational step using scalar-relativistic eigenfunctions as basis. The surface was simulated by placing a slab of six quintuple layers in vacuum. A grid of 35$\times$35$\times$1 points were used in the calculations, equivalent to 120 k-points in the irreducible BZ and 2450 k-points in the first BZ.

The simplest signature of topological Z2 order is the existence of a single spin-polarized Dirac cone on its surface, with the Dirac node located at a momentum \textbf{k}$_T$ in the surface Brillouin zone (BZ), where \textbf{k}$_T$ satisfies \textbf{k}$_T$ = -\textbf{k}$_T$ + \textbf{G}, \textbf{G} being a surface reciprocal lattice vector \cite{Fu:STI2}. Because this node is protected by time-reversal-symmetry \cite{Fu:STI2}, the spin polarization is guaranteed to rotate by 360$^{\circ}$ about its Fermi surface \cite{Hsieh_Science} (Fig.~\ref{fig:STI}(b)). Figure~\ref{fig:STI}(c) shows a spin-integrated ARPES intensity map of Bi$_2$Te$_3$ (111) taken at a constant energy $E_F$, which spans several surface BZs. The density of states at $E_F$ is clearly localized only to a small area around the \textbf{k}$_T$  = $\bar{\Gamma}$ points, suggesting a very small concentration of surface carriers. A high resolution intensity map around
$\bar{\Gamma}$ shows that the density of states is distributed about a ring enclosing $\bar{\Gamma}$ (Fig.~\ref{fig:STI}(d)). This allows for a calculation of the surface carrier density $n$ using the Luttinger
theorem $n = \pi k_F^2/(A_{BZ}A_{R})$, where $k_F$ is the Fermi momentum, $A_{BZ}$ is the area of the surface BZ and $A_R$ is the area of the surface unit cell, which yields $n \sim 1.7 \times 10^{12}$ cm$^{-2}$. The dispersion of the Dirac cone (Figure~\ref{fig:STI}(e)), imaged an hour after cleavage, shows a Dirac node located at a binding energy ($E_B$) of around -130 meV. Because experimental proposals to measure non-Maxwell electrodynamics \cite{Franz,Qi_monopole,Zhang_STM} or non-Abelian quasiparticles \cite{Fu_Majorana2,Akhmerov} require the surface state to be made insulating by lifting the spin degeneracy of the Dirac node with a magnetic field, a Dirac node close to $E_F$ means that this can be achieved using smaller magnetic fields.

\begin{figure}
\includegraphics[scale=0.55,clip=true, viewport=0.0in 0in 12.5in 8.2in]{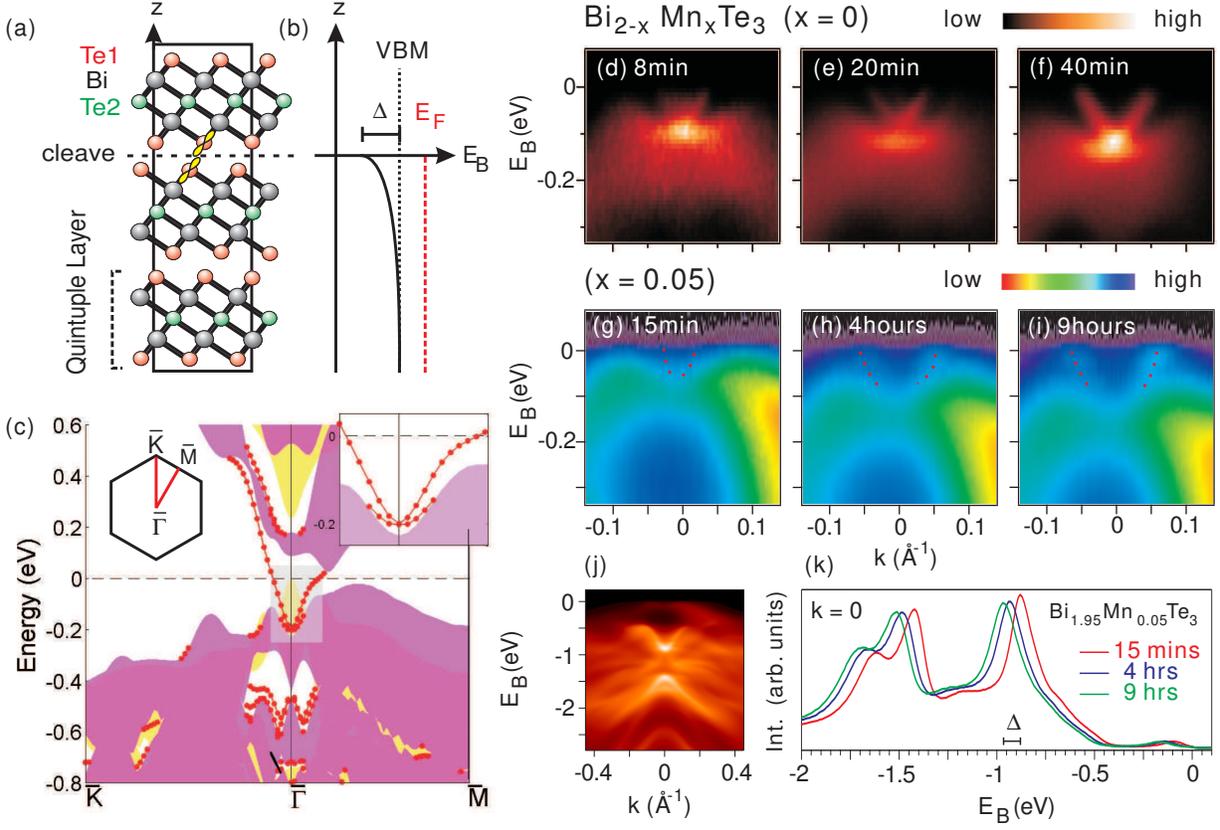}
\caption{\label{fig:Time}. \textbf{Slow dynamics of the surface Dirac cone dispersion.} (a) The crystal structure of Bi$_2$Te$_3$ viewed parallel to the quintuple layers. The Te(1) 5$p$ orbitals that form the inter-quintuple layer van der Waals bonds are shown in yellow. (b) Schematic of the band bending of the bulk valence band maximum (VBM) near the cleaved surface. (c) Calculated band structure along the \={K}-$\bar{\Gamma}$-\={M} cut of the Bi$_2$Te$_3$(111) BZ. Bulk band projections are represented by the shaded areas. The band structure results with spin-orbit coupling (SOC) are presented in purple and that without SOC in yellow. No pure surface band is observed within the bulk band gap without SOC (black lines). One pure gapless surface band crossing $E_F$ is observed when SOC is included (red lines). Inset shows enlargement of low energy region near $\bar{\Gamma}$. Spin-integrated ARPES spectra of Bi$_2$Te$_3$ along the $\bar{\Gamma}$-\={M} direction taken with $h\nu$ = 30 eV (d) 8 mins, (e) 20 mins, and (f) 40 mins after cleavage in UHV. Analogous ARPES spectra for Bi$_{1.95}$Mn$_{0.05}$Te$_3$ (g) 15 mins, (h) 4 hours and (i) 9 hours after cleavage, showing a slower relaxation rate. Red lines are guides to the eye. (j) High energy valence band ARPES scan of Bi$_{1.95}$Mn$_{0.05}$Te$_3$ and (k) the energy distribution curves at $\bar{\Gamma}$ taken at different times.}
\end{figure}

ARPES spectra taken as a function of time show that the binding energy of the Bi$_2$Te$_3$ surface Dirac node exhibits a pronounced time dependence, increasing from $E_B \sim$ -100 meV 8 minutes after cleavage to $E_B \sim$ -130 meV at 40 minutes (Fig.~\ref{fig:Time}(d)-(f)), in agreement with a previous report \cite{Noh}. Such behavior has been attributed to a downward band bending near the surface (Fig.~\ref{fig:Time}(b)) that is caused by the breaking of inter-quintuple layer van der Waals Te(1)-Te(1) bonds (Fig.~\ref{fig:Time}(a)) and the formation of Te(1)-Bi bonds \cite{Urazhdin, Mishra}, which increases the electron density near the surface. To gain insight into the surface state formation process, we carried out $ab$ $initio$ calculations with optimized lattice parameters. The calculations (Fig.~\ref{fig:Time}(c)) show that the surface Dirac cone only appears when spin-orbit coupling is switched on and that its dispersion closely matches the late-time ARPES spectrum (Fig.~\ref{fig:Time}(f)). The calculated position of the Dirac node lies in the bulk band gap, which corroborates our experimental data showing a strong intensity near the Dirac node and a drastic weakening away from $\bar{\Gamma}$ as the surface band merges with the bulk bands to form resonance states \cite{Hufner,Hsieh_Nature,Hsieh_Science}. The slow dynamics of the band bending process suggests that the local surface lattice structure is coupled to the local surface charge density. Equilibration involving both atomic and charge redistribution may thus be significantly delayed by structural defects such as site inversions or vacancies. By systematically increasing the defect concentration through Mn/Bi substitution, we show that band bending can be slowed by up to 10 fold (Fig.~\ref{fig:Time}(g)-(i)), allowing a wider range of its intrinsic relaxation time scale to be accessed. ARPES valence band spectra (Fig.~\ref{fig:Time}(k)) of Bi$_{1.95}$Mn$_{0.05}$Te$_3$ taken over a 15 hour period show that the positions of the valence band edges shift downward by a total energy of around 100 meV (Fig.~\ref{fig:Time}(k)), representing the total magnitude of band bending $\Delta$.

\begin{figure}
\includegraphics[scale=0.5,clip=true, viewport=-0.4in 0in 8.5in 11.0in]{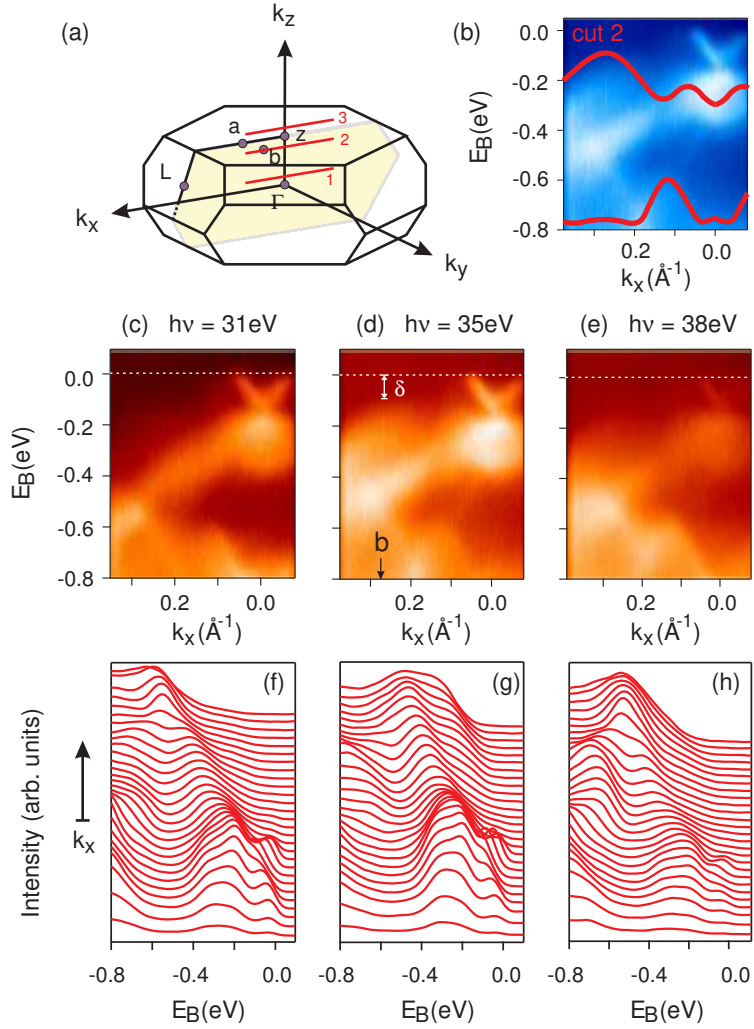}
\caption{\label{fig:Bulk}. \textbf{Observation of insulating bulk-like states supporting a six-peak electronic structure.} (a) Bulk rhombohedral Brillouin zone of Bi$_2$Te$_3$. According to LDA band structure calculations \cite{Youn}, six valence band maxima are located at the \textbf{b} points that are related to one another by 60$^{\circ}$ rotations about $\hat{z}$. The red lines show the momentum space trajectories of the ARPES scans taken using $h\nu$ = 31 eV, $h\nu$ = 35 eV and $h\nu$ = 38 eV. (b) Calculated band structure through the \textbf{b} point along cut 2 that has been rigidly shifted to match the superimposed corresponding ARPES cut to account for the Fermi level shift in real materials. (c) to (e) show ARPES intensity maps along the $h\nu$ = 31 eV, 35 eV and 38 eV trajectories respectively, obtained one hour after sample cleavage. The in-plane momentum component of the \textbf{b} point is marked by an arrow in (d), and the energy of the valence band maximum relative to $E_F$ is denoted $\delta$. (f) to (g) show the energy distribution curves corresponding to images (c) to (e) respectively.}
\end{figure}

Our theoretical calculations show that the bulk valence band maxima (VBM) in Bi$_2$Te$_3$ lie in the $\Gamma Z L$ planes of the bulk BZ around 15 meV below $E_F$ (Fig.~\ref{fig:Bulk}(b)), giving rise to VBM in each of the six mirror planes in agreement with theoretical proposals \cite{Youn,Mishra}. In order to measure the locations of these bulk VBM in the three-dimensional BZ, we performed a series of ARPES scans along the trajectories shown by the red lines in Figure~\ref{fig:Bulk}(a) with varying incident photon energies to access multiple values of $k_z$. In order to identify subtle changes in the band structure with changing $k_z$, it is desirable to work with a band structure that is stable in time. Therefore all $hv$-dependent scans were taken more than an hour after cleavage. However, because ARPES is only sensitive to the topmost quintuple layer at the photon energies we use \cite{Hufner}, the three-dimensional band structure we measure is only bulk-like in the sense that the energy positions may differ from the true bulk states due to near-surface band bending effects. Figures~\ref{fig:Bulk}(c)-(e) show a series of ARPES intensity images of the band dispersions along momentum space trajectories in the $k_x$-$k_z$ plane taken using photon energies of 31 eV, 35 eV and 38 eV respectively (Fig.~\ref{fig:Bulk}). The Dirac cone near $E_F$ centered at $k_x$ = 0 \AA$^{-1}$ shows no dispersion with photon energy, supporting its surface state origin. In contrast, a strongly $k_z$ dispersive hole-like band is observed near $k_x$ = 0.27 \AA$^{-1}$, whose peak is found to lie closest to $E_F$ at $h\nu$ = 35 eV (Fig.~\ref{fig:Bulk}(d)). Using the free electron final state approximation \cite{Hufner, Noh}, this VBM lies close to the (0.27, 0, 0.27) point in the bulk BZ, which agrees well with our calculations. The energy position $\delta$ of the VBM lies approximately 100 meV below $E_F$ based on its peak position in the energy distribution curves (Fig.~\ref{fig:Bulk}(g)), which is lower than that found in our calculations by $\sim$85 meV. This discrepancy agrees well with magnitude of the near-surface band bending $\Delta$ that is also responsible for the lowering of the surface band energies (Fig.~\ref{fig:Time}). We conclude that the three-dimensional states near the surface region behave as a large band gap insulator. 

\begin{figure}
\includegraphics[scale=0.6,clip=true, viewport=-0.1in 0in 9.0in 7.0in]{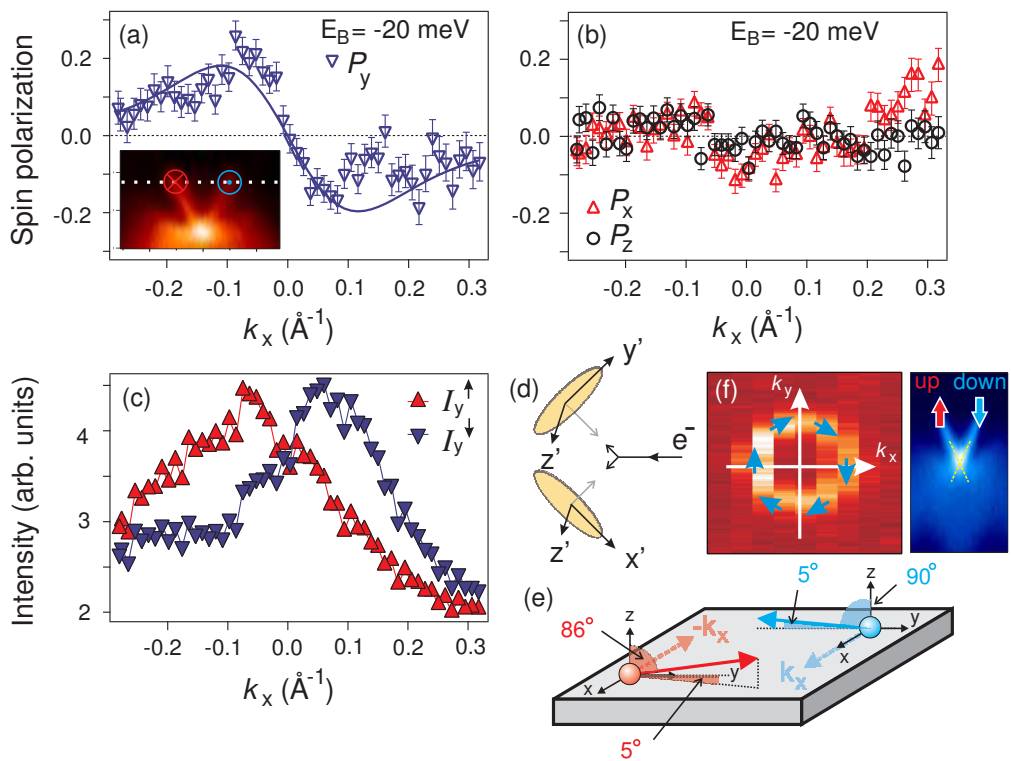}
\caption{\label{fig:SpinBi2Te3}. \textbf{Helicity of surface Dirac
cone with a non-trivial topology}. (a) Measured $y$ component of spin-polarization along the $\bar{\Gamma}$-\={M} direction at $E_B$ = -20 meV, which only cuts through the surface states. Inset shows energy position of this cut together with fitted spin-polarization directions. The solid line is a fit based on the procedure of Meier $et$ $al$ \cite{Meier}. (b) Measured $x$ (orange diamonds) and $z$ (green circles) components of spin-polarization along the $\bar{\Gamma}$-\={M} direction at $E_B$ = -20 meV. (c) Spin-resolved spectra obtained from $y$ component spin polarization data. (d) Geometry of the Mott detector axes. (e) Fitted values of the spin polarization vector demonstrating helicity of the Dirac cone. The largest source of uncertainty in the fitted angles comes from the time dependence of the surface state spectrum. (f) Spin-integrated ARPES intensity map of the surface state Fermi surface with its the measured helicity superimposed. This originates from the spin-polarized Dirac cone shown to its right.}
\end{figure}

To determine the helicity of the surface Dirac cones, we performed spin-resolved ARPES scans \cite{GayDunning} using a double Mott detector configuration (Fig.~\ref{fig:SpinBi2Te3}(d)), which allows all three components of the spin vector of a photoelectron to be measured \cite{Meier}. We analyzed the spin-polarization of photoelectrons emitted at $E_B$ = -20 meV along momentum space cuts traversing the surface states (Fig.~\ref{fig:SpinBi2Te3}). Although scans were taken hours after cleavage in vacuum to allow time for the surface states to stabilize, the surface band dispersion continued to slowly evolve over the course of the scans. In effort to compromise between the effects of changes in the surface state dispersion with time, increasing surface contamination over time, and the typical hours-long counting times needed for spin-resolved measurements, we collected data with a level of statistics sufficient to show the spin-polarized character of the surface states.
Figures~\ref{fig:SpinBi2Te3}(a) and (b) show the spin polarization spectra of the $x$, $y$ and $z$ (out-of-plane) components along the $\bar{\Gamma}$-\={M} direction. In the $x$ and $z$ directions, no clear signal can be discerned within the margins of statistical error. In the $y$ direction on the other hand, clear polarization signals of equal magnitude and opposite sign are observed for the surface bands on either side of $\bar{\Gamma}$, evidence that they form a time-reversal-invariant spin-orbit split pair. Because $\bar{\Gamma}$-\={M} is along a line of mirror symmetry in the Bi$_2$Te$_3$ crystal, it is expected that the spins are constrained along the $\pm\hat{y}$ direction along this cut. A fit based on the two-step routine developed by Meier $et$ $al.$ \cite{Meier} is consistent with 100\% polarized spins that are directed completely along the $\pm\hat{y}$ direction as shown in Figure~\ref{fig:SpinBi2Te3}(e). Figure ~\ref{fig:SpinBi2Te3}(c) shows the spin-resolved spectra ($I_y^{\uparrow,\downarrow}$) calculated from $P_y$ according to $I_y^{\uparrow} = I_{tot}(1+P_y)/2$ and $I_y^{\downarrow} = I_{tot}(1-P_y)/2$, where $I_{tot}$ is the spin-averaged intensity, which shows a peak to peak separation in agreement with 2$k_F$. The non-Lorentzian lineshape of the $I_y^{\uparrow}$ and $I_y^{\downarrow}$ curves and their non-exact merger at large $|k_x|$ is likely due to the time evolution of the surface band dispersion.

Our combination of conventional and spin-resolved ARPES results show that Bi$_{2-x}$Mn$_x$Te$_3$ hosts a single unpaired helical Dirac cone on its (111) surface that is fully spin-polarized, opening rare opportunities for Dirac spin materials research beyond the chiral variety for the first time. Our finding of large-gap insulating behavior of the near surface bulk-like spin-degenerate states in the $x$ = 0 compound, together with a low lying spin-resolved Dirac node makes Bi$_2$Te$_3$ suitable for studying topological effects via interface transport.



\newpage

\begin{thebibliography}{99}

\bibitem{Novoselov(Graphene)}
K.S. Novoselov $et$ $al.$ Two-dimensional gas of massless Dirac fermions in graphene. \textit{Nature} \textbf{438}, 197-200 (2005).

\bibitem{Checkelsky}
J.G. Checkelsky, L. Li and N.P. Ong. Divergent resistance of the Dirac point in graphene: Evidence for a transition in high magnetic field. \textit{Phys. Rev. B} \textbf{79}, 115434 (2009).

\bibitem{Li}
L. Li $et$ $al.$ Phase transitions of Dirac electrons in Bismuth. \textit{Science} \textbf{321}, 547-550 (2008).

\bibitem{Fuseya}
Y. Fuseya, M. Ogata and H. Fukuyama. Interband contributions from the magnetic field on Hall effects for Dirac electrons in Bismuth. \textit{Phys. Rev. Lett.} \textbf{102}, 066601 (2009).

\bibitem{Day}
C. Day. Exotic spin textures show up in diverse materials. \textit{Physics Today} \textbf{62}, 12-13 (2009).

\bibitem{Koralek}
J.D. Koralek $et$ $al.$ Emergence of the persistent spin helix in semiconductor quantum wells. \textit{Nature} \textbf{458}, 610-613 (2009).

\bibitem{Fu:STI2}
L. Fu, C.L. Kane and E.J. Mele. Topological insulators in three dimensions. \textit{Phys. Rev. Lett.} \textbf{98}, 106803 (2007).

\bibitem{Ran_dislocation}
Y. Ran, Y. Zhang and A. Vishwanath. One-dimensional topologically protected modes in topological insulators with lattice dislocations. \textit{Nature Phys.} \textbf{5}, 298-303 (2009).

\bibitem{Essin}
A. Essin, J.E. Moore and D. Vanderbilt. Magnetoelectric polarizability and axion electrodynamics in crystalline insulators. Preprint at $<$http://arxiv.org/abs/0810.2998$>$ (2008).

\bibitem{Franz}
M. Franz. High-energy physics in a new guise. \textit{Physics} \textbf{1}, 36 (2008).

\bibitem{Qi_monopole}
X.-L. Qi, R. Li, J. Zang and S.-C. Zhang. Inducing a magnetic monopole with topological surface states. \textit{Science} \textbf{323}, 1184-1187 (2009).

\bibitem{zhang_bete}
H. Zhang $et$ $al.$  Topological Insulators at Room Temperature. Preprint at arXiv:0812.1622v1 (2008).

\bibitem{Moore_exciton}
B. Seradjeh, J.E. Moore and M. Franz. Exciton condensation and charge fractionalization in a topological insulator film. Preprint at $<$http://arxiv.org/abs/0902.1147$>$ (2009).

\bibitem{Fu_Majorana2}
L. Fu and C.L. Kane. Probing neutral Majorana fermion edge modes with charge transport. Preprint at $<$http://arxiv.org/abs/0903.2427$>$ (2009).

\bibitem{Akhmerov}
A.R. Akhmerov, J. Nilsson and C.W.J. Beenakker. Electrically detected interferometry of Majorana fermions in a topological insulator. Preprint at $<$http://arxiv.org/abs/0903.2196$>$ (2009).

\bibitem{Moore:STI1}
J.E. Moore and L. Balents. Topological invariants of time-reversal-invariant band structures. \textit{Phys. Rev. B} \textbf{75} 121306(R) (2007).

\bibitem{Hsieh_Nature}
D. Hsieh $et$ $al.$ A topological Dirac insulator in a quantum spin Hall phase. \textit{Nature} \textbf{452}, 970-974 (2008).

\bibitem{Xia}
Y. Xia $et$ $al$. Electrons on the surface of Bi$_2$Se$_3$ form a topologically-ordered two dimensional gas with a non-trivial Berry's phase. \textit{Nature Phys.} in press (2009).

\bibitem{Noh}
H.-J. Noh $et$ $al.$ Spin-orbit interaction effect in the electronic structure of {Bi$_2$Te$_3$} observed by angle-resolved photoemission spectroscopy. \textit{Europhys. Lett.} \textbf{81}, 57006 (2008).

\bibitem{Hor}
Y.S. Hor $et$ $al.$ p-type Bi$_{2}$Se$_3$ for topological insulator and thermoelectric applications. Preprint at $<$http://arxiv.org/abs/0903.4406$>$ (2009).

\bibitem{Schnyder}
A.P. Schnyder, S. Ryu, A. Furusaki and A.W.W. Ludwig. Classification of topological insulators and superconductors in three spatial dimensions. \textit{Phys. Rev. B} \textbf{78} 195125 (2008).


\bibitem{HoeschFSmap}
M. Hoesch $et$ $al.$ Spin-polarized Fermi surface mapping. \textit{J. Elect. Spec. Rel. Phen.} \textbf{124}, 263-279 (2002).

\bibitem{Wien}
P. Blaha $et$ $al.$ Computer code WIEN2K. Vienna University of Technology, Vienna (2001).

\bibitem{Ernzerhof}
J.P. Perdew, K. Burke and M. Ernzerhof. Generalized gradient approximation made simple. \textit{Phys. Rev. Lett.} \textbf{77}, 3865-3868 (1996).

\bibitem{Hsieh_Science}
D. Hsieh $et$ $al.$ Observation of unconventional quantum spin textures in topological insulators. \textit{Science} \textbf{323}, 919-922 (2009).

\bibitem{Zhang_STM}
Q. Liu, C.-X Liu, C. Xu, X.-L. Qi and S.-C. Zhang. Magnetic impurities on the surface of a topological insulator. Preprint at $<$http://arxiv.org/abs/0808.2224$>$ (2008).

\bibitem{Urazhdin}
S. Urazhdin $et$ $al.$ Surface effects in layered semiconductors Bi$_2$Se$_3$ and Bi$_2$Te$_3$. \textit{Phys. Rev. B} \textbf{69}, 085313 (2004).

\bibitem{Mishra}
S.K. Mishra, S. Satpathy and O. Jepsen. Electronic structure and thermoelectric properties of bismuth telluride and bismuth selenide. \textit{J. Phys: Condens. Mat.} \textbf{9}, 461-470 (1997).

\bibitem{Hufner}
S. H{\"{u}}fner. Photoelectron Spectroscopy. Springer, Berlin (1995).

\bibitem{Youn}
S.J. Youn and A.J. Freeman. First-principles electronic structure and its relation to the thermoelectric properties of Bi$_2$Te$_3$. \textit{Phys. Rev. B} \textbf{63}, 085112 (2001).

\bibitem{GayDunning}
T.J. Gay and F.B. Dunning. Mott electron polarimetry. \textit{Rev. Sci. Instr.} \textbf{63}, 1635-1651 (1991).

\bibitem{Meier}
F. Meier, H. Dil, J. Lobo-Checa, L. Patthey and J. Osterwalder. Quantitative vectorial spin analysis in angle-resolved photoemission: Bi/Ag(111) and Pb/Ag(111). \textit{Phys. Rev. B} \textbf{77}, 165431 (2008).

\end{thebibliography}

\newpage

\end{document}